\documentclass[twoside,a4paper,11pt]{sca}
\def\aj{AJ}%
\def\araa{ARA\&A}%
\def\apj{ApJ}%
\def\apjl{ApJ}%
\def\aap{A\&A}%
\def\mnras{MNRAS}%

\def\age{t}

\def\gyr{{\rm Gyr}}
\def\reff{r_{\rm eff}}
\def\rh{r_{\rm h}}
\def\rhn{r_{\rm h0}}
\def\msun{{\rm M}_{\odot}}

\def\myr{{\rm Myr}}
\def\pc{{\rm pc}}
\def\rhoh{\rho_{\rm h}}
\def\tcr{t_{\rm cr}}

\def\trh{t_{\rm rh}}
\def\trhn{t_{\rm rh0}}

\usepackage{graphicx,latexsym,amssymb}
\usepackage{hyperref}
\usepackage{movie15}
\usepackage{natbib, natbibspacing}  

\begin{document}

\pagenumbering{arabic}
\pagestyle{myheadings}
\thispagestyle{empty}
{\flushright\includegraphics[width=\textwidth,bb=90 650 520 700]{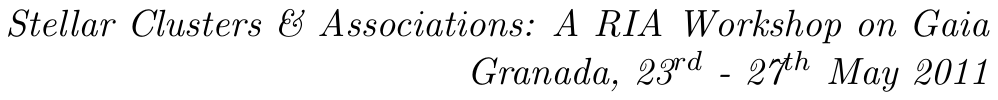}}
\vspace*{0.2cm}
\begin{flushleft}
{\bf {\LARGE
Dynamical evolution of stellar clusters
%
}\\
\vspace*{1cm}
Mark Gieles
%
}\\
\vspace*{0.5cm}
Institute of Astronomy,\,University of Cambridge,\,Madingley\,Road,\,Cambridge, CB3\,0HA,\,UK\\
%
\end{flushleft}
%
\markboth{
Dynamical evolution of stellar clusters
}{ 
Mark~Gieles
%
}
\thispagestyle{empty}
\vspace*{0.4cm}
\begin{minipage}[l]{0.09\textwidth}
\ 
\end{minipage}
\begin{minipage}[r]{0.9\textwidth}
\vspace{1cm}
\section*{Abstract}{\small
The evolution of star clusters is determined by several internal and
external processes. Here we focus on two dominant internal effects,
namely energy exchange between stars through close encounters
(two-body relaxation) and mass-loss of the member stars through
stellar winds and supernovae explosions.  Despite the fact that the
former operates on the relaxation timescale of the cluster and the
latter on a stellar evolution timescale, these processes work together
in driving a nearly self-similar expansion, without forming (hard)
binaries. Low-mass clusters expand more, such that after some time the
radii of clusters depend very little on their masses, even if all
clusters have the same (surface) density initially.  Throughout it is
assumed that star clusters are in virial equilibrium and well within
their tidal boundary shortly after formation, motivated by
observations of young ($\sim$few~Myrs) clusters. We start with a
discussion on how star clusters can be distinguished from (unbound)
associations at these young ages.
\normalsize}
\end{minipage}

\section{Introduction }
\label{sec:intro}
\citet{amb47} introduced the term association in reference to loose
agglomerates and he pointed out in subsequent studies that it is
unlikely that they are bound by their own gravity \citep[see
  also][]{1964ARA&A...2..213B}.  It is often stated that the majority
of stars form in star clusters and that there is a high rate of early
cluster disruption \citep[e.g.][]{2003ARA&A..41...57L}. In this view
associations are clusters that have expanded. But if the star
formation process is hierarchical then only a small fraction of the
newborn stars reside in agglomerates that satisfy the conditions
necessary to be bound by self-gravity at formation
\citep[e.g.][]{2005ApJ...623..650K,2008ApJ...672.1006E,
  2010MNRAS.409L..54B}. When observational samples of star clusters
are used to support either one of the above scenarios it is vital to
know how star clusters are separated from associations \citep{B11}.
Here we provide a definition of the distinction between these two
classes of stellar agglomerates (\S~\ref{sec1}).  In \S~\ref{sec2} we
present results of $N$-body simulations of star clusters including the
effect of stellar evolution.

\section{The distinction between star clusters and associations}
\label{sec1}
To illustrate our case we use the recent literature compilation of
young massive clusters and associations of \citet[][hereafter
  PZMG10]{2010ARA&A..48..431P}.  This sample consists of stellar
agglomerates for which a value of the half-light radius $\reff$, mass
$M$, and age $\age$ are available in literature. The sample contains 105
agglomerates with $M\gtrsim10^4\,\msun$ and $\age\lesssim100\,\myr$ in
nearby ($\lesssim10\,$Mpc) galaxies, including the Milky Way.  The
ratio of the age over the crossing time, $\Pi=\age/\tcr$, can be
used to separate star clusters from associations \citep[PZMG10;
][]{2011MNRAS.410L...6G}.  Objects that are older than their crossing
time ($\Pi>1$) are most likely bound star clusters, whereas objects
with $\Pi<1$ are expected to be unbound associations.  The crossing
time is defined in terms of empirical cluster parameters
\begin{equation}
\tcr\equiv10\left(\frac{\reff^3}{GM}\right)^{1/2},
\label{eq:tcr}
\end{equation}
where $G\simeq4.5\times10^{-3}\,\pc^{3}\,\msun^{-1}\,\myr^{-2}$ is the
gravitational constant. Equation~(\ref{eq:tcr}) is valid for systems
in virial equilibrium, because the formal definition includes the
root-mean square velocity dispersion: $\tcr\propto\reff/\sigma$ and in
virial equilibrium $\sigma\propto\sqrt{M/\reff}$. However, $\sigma$
values are available for fewer agglomerates and at young ages the
empirically determined $\sigma$ can be higher because of orbital
motions of multiples \citep{2010MNRAS.402.1750G}. Surprisingly, the
more convenient equation~(\ref{eq:tcr}) fascilites in making the
distinction between bound and unbound objects. Because super-virial
associations expand with a (roughly) constant velocity, their crossing
evolves as $\tcr\propto\reff\propto t$. Using equation~(\ref{eq:tcr})
for super-virial objects, therefore, overestimates $\tcr$ and underestimates $\Pi$, thereby
pushing them more into the unbound regime.

\begin{figure}[!t]
\center
\includegraphics[width=0.8\textwidth]{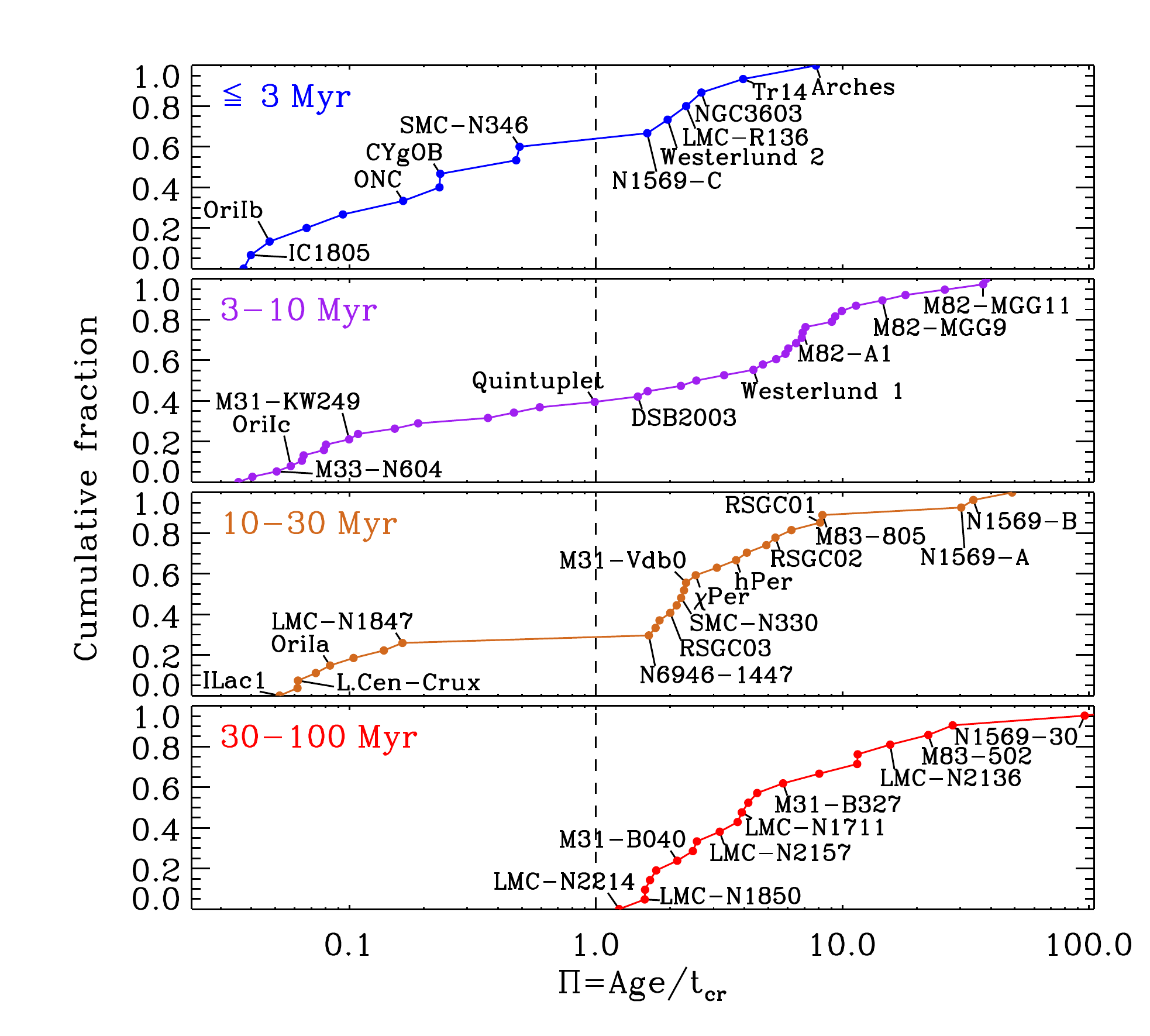} 
\caption{\label{fig1} Cumulative distribution of $\Pi$ values for
  nearby ($\lesssim10\,$Mpc) agglomerates in different age bins. The
  vertical dashed line indicates the value $\Pi=1$ that separates star
  clusters from associations. From \citet{2011MNRAS.410L...6G}.  }
\end{figure}

In Fig.~\ref{fig1} we show the cumulative distribution of $\Pi$ values
of all objects in different age bins.  The top panel shows that the
youngest age bin is a continuous distribution from associations with
$\Pi\sim0.03$ to star clusters with $\Pi\sim10$.  There seems not to
be a distinct mode of star cluster formation, but rather a smooth
transition from star clusters to
associations. \citet{2010MNRAS.409L..54B} come to a similar conclusion
based on the surface density distribution of young stellar objects in
the solar neighbourhood

The bottom panel shows that the oldest agglomerates all have
$\Pi\gtrsim1$ and these are, therefore, star clusters.  The
intermediate age curves contain both associations and star clusters.
If we interpret the curves for the different age bins as an
evolutionary sequence then a distinct gap develops between star
clusters and associations around $\sim10\,$Myr at a value of
$\Pi\simeq1$.  At older ages an observer should be able to make an
unambiguous distinction between an (unbound) association and a (bound)
star cluster using this straight-forward method.

For the youngest (continuous) distribution a useful first order separation can still be made at a value of $\Pi=1$, as
can be noted from the labels of several well known star clusters and
associations.  Independent confirmation comes from recent
determinations of velocity dispersions of (resolved) young
($\sim1-2\,\myr$) star clusters. For NGC~3603
\citep{2010ApJ...716L..90R}, Westerlund~1 \citep[][and M.~Cottaar in
  this volume]{2007A&A...466..151M} and R136 in 30~Doradus
\citep[][H{\'e}nault-Brunet in prep.]{2009AJ....137.3437B} it was
found that the dynamical mass estimates agree very well with the
photometric masses, suggesting that these clusters are in virial
equilibrium and stable (i.e. bound). This is also what their $\Pi$
values suggest (see Fig.~\ref{fig1}). In the next section we consider
the dynamical evolution of such star clusters.

\section{$N$-body simulations of star clusters}
\label{sec2}
We want to understand the evolution of a stellar cluster with a
realistic stellar mass function in which the stars evolve and lose
mass in time \citep[][hereafter G10]{2010MNRAS.408L..16G}. Because
most young star clusters are very deeply embedded in their tidal limit
(see e.g. \citealt{2008ApJ...675.1319H} and
\citealt{2011MNRAS.412.2469G} for the cases of NGC~3603 and
Westerlund~1, respectively) we ignore the effect of a tidal
cut-off. The transition from expansion-dominated evolution to tidally
limited evolution is considered  in more detail elsewhere
\citep{2011MNRAS.413.2509G}.

We model clusters with initial half-mass relaxation timescales ranging
from $\trhn\simeq1\,\myr$ to $\trhn\simeq4\,\gyr$ by considering
different combinations of $[N,\log\rhoh]$ at the start of the
simulation.  The stars follow a Kroupa~(2001) initial mass function
between $0.1\,\msun$ and $100\,\msun$. We use the \texttt{kira}
integrator and the stellar evolution package \texttt{SeBa} for solar
metallicity \citep{2001MNRAS.321..199P}.  The retention fraction of
black holes and neutron stars was set to zero.

\begin{figure}[!t]
\center \includegraphics[width=0.65\textwidth]{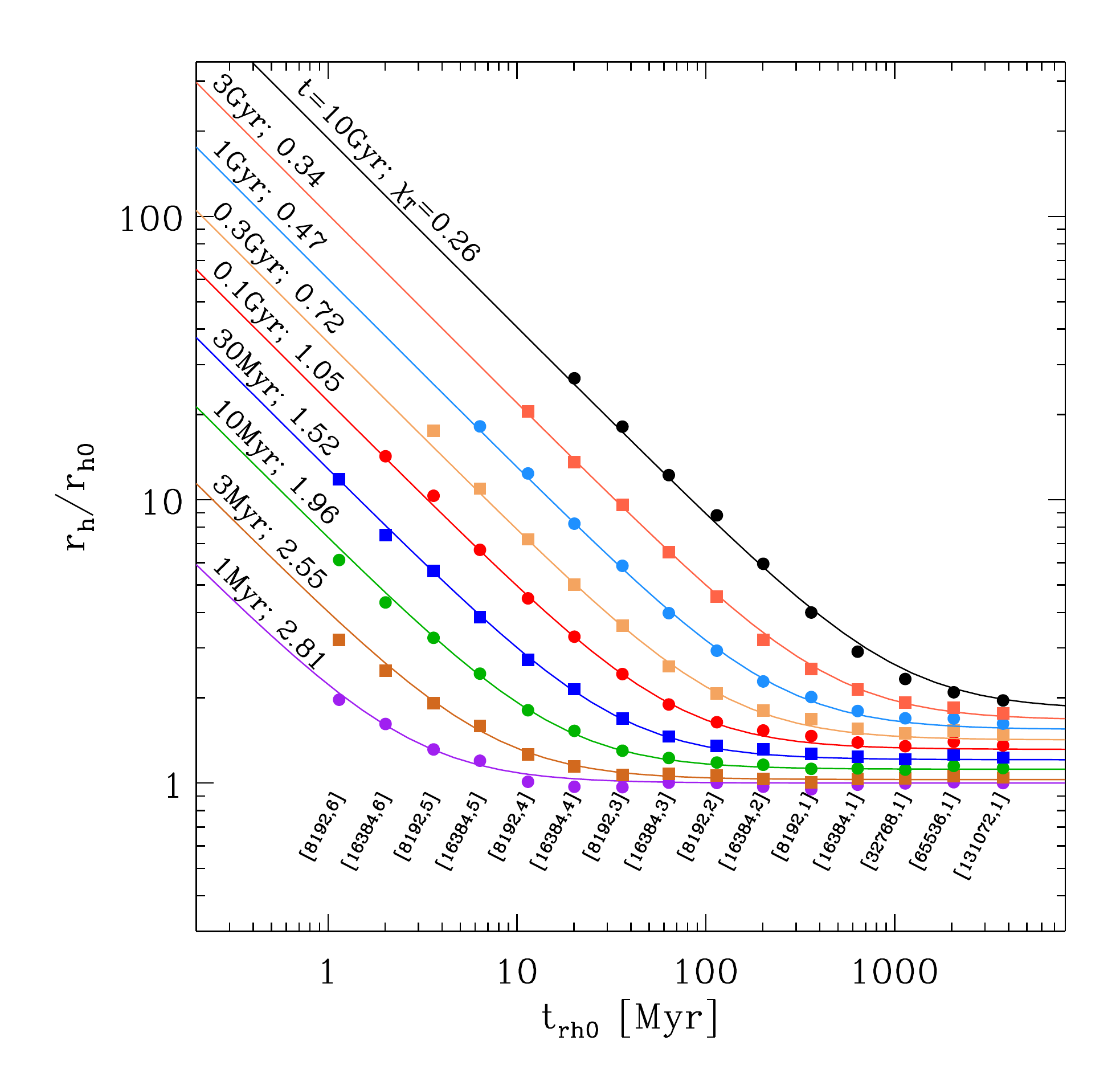}
\caption{\label{fig2} Expansion from the $N$-body runs including the
  effect of stellar evolution. The full lines show a functional
  description that is used to evolve the inital mass-radius relation
  in Figs.~\ref{fig3} and \ref{fig4} (see G10 for more details).  }
\end{figure}

In Fig.~\ref{fig2} we show how $\rh/\rhn$ increases with time as a
function of $\trhn$. The asymptotic behaviour of these results can
easily be understood by considering the extremes. Clusters that are
dynamically young (low $\age/\trhn$) expand adiabatically in order to
retain virial equilibrium after stellar mass loss.  This adiabatic
expansion is slow in time and gives a maximum increase of
$\rh/\rhn\simeq2$ after a Hubble time.  At the other extreme we have
clusters that are dynamically old (high $\age/\trhn$) and they expand
in a way that is comparable to what is found for equal-mass clusters
in the sense that all clusters evolve towards the same $\trh$.

We conclude that the energy that is released to the core of the
cluster as a result of stellar mass-loss acts as a central energy
source that is in `balance' with the energy that is driven outwards by
2-body relaxation, without the need of forming binaries in the
core. One of the consequences of this interplay between stellar
evolution and dynamical relaxation is that there is no sharp
transition between a stellar evolution dominated phase and a
relaxation dominated phase.

\section{Comparing the theory to  observations}
\label{sec:application}

With the expression for the evolution of the radius as a function of
$\trhn$ at hand (equation~6 in G10) we can easily calculate the
evolution of any initial mass-radius relation. We first apply the
results to the sample of young objects discussed in the previous
section.

\begin{figure}[!t]
\center
\includegraphics[width=\textwidth]{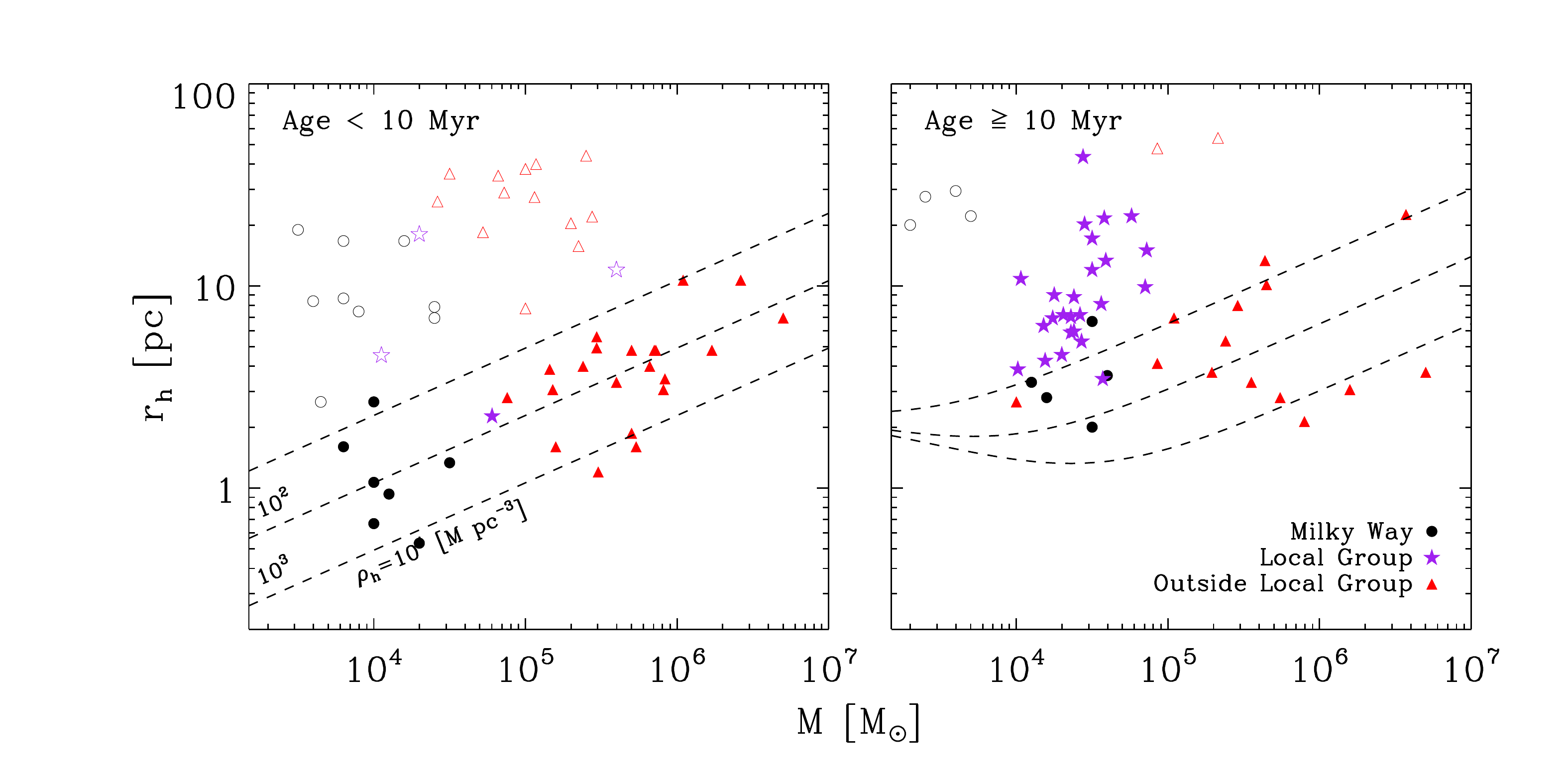} 

\caption{\label{fig3} Mass-radius relation for all clusters (filled
  symbols) and associations (open symbols) from PZMG10, using $\rh=
  (4/3)\reff$. Lines of constant half-mass density, $\rhoh \equiv 3 M
  /(8\pi \rh^3)$, are over-plotted. The clusters are subdivided into
  two groups: younger than 10\,Myr (left) and between 10\,Myr and
  100\,Myr (right). There is a lack of compact ($\lesssim2\,\pc$) and low-mass ($\lesssim10^4\,\msun$) clusters in the older bin
  due to dynamical expansion.  }
\end{figure}

In Fig.~\ref{fig3} we show mass-radius diagrams for all objects for
two age ranges: $< 10\,\myr$ (left panel) and $10-100\,\myr$ (right
panel). The dashed lines in the left panel are lines of constant
(arbitrary) $\rhoh$. In the right panel we have evolved these lines to
an age of 100\,\myr. These lines are lower limits to the radius at the
left-side of this diagram. There are indeed no compact
($\lesssim2\,\pc$) low-mass ($\lesssim10^4\,\msun$) clusters in this age bin.

\begin{figure}[!t]
\center
\includegraphics[width=0.6\textwidth]{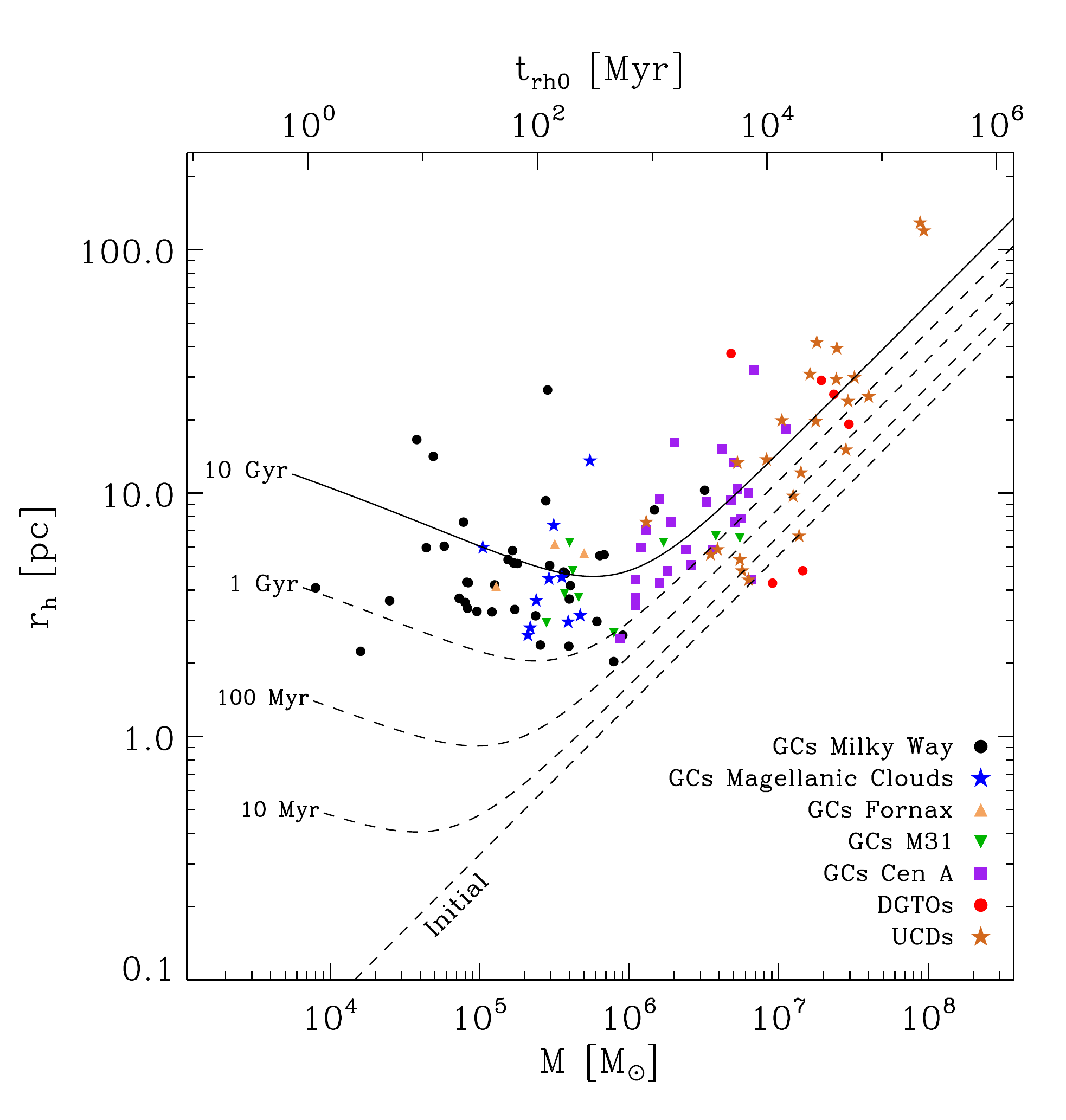} 
\caption{\label{fig4} Mass-radius values for hot stellar systems (see
  G10 for more details on the data).  The lines show the evolution of
  the mass-radius relation. The break at $\sim10^6\,\msun$ at
  $10\,\gyr$ is because lower mass objects have expanded.  }
\end{figure}

Next we apply our result to the mass-radius relation of old stellar
systems in the mass range $\sim10^4-10^8\,\msun$.  In Fig.~\ref{fig4}
we show how a Faber-Jackson type initial mass-radius relation
\citep{2005ApJ...627..203H} evolves in time together with data points
of (old) globular clusters and ultra-compact dwarfs (UCDs) in different galaxies that cover
the mass regime we are interested in. For high $t/\trhn$ the radius is
set by $M_0$, independent of $\rhn$, while for low $t/\trhn$ we are
seeing roughly the initial mass-radius relation. 
At an age of $10\,\gyr$ the break
between the two regimes occurs at $M_0\simeq 10^6\,\msun$ and
at that age systems with this mass have $\trh/t\simeq0.8$.
\citet{2008A&A...487..921M} noticed already that the break occurs at
systems with $\trh$ roughly equal to a Hubble time. Here we give a
quantitative explanation for it.

We conclude that the evolution of almost all (old) globular clusters
is `balanced', in the sense that the rate of central energy production
equals the flow of energy due to 2-body relaxation\footnote{We
  deliberately avoid the term `post-collapse' evolution because
  core-collapse is not required for the evolution to be `balanced',
  see the discussion in \citet{2011MNRAS.413.2509G}}.  An important
property of this `balanced' evolution is that the half-mass
radius is independent of its initial value and is a function of
the number of stars and the age only. It is therefore not possible to
infer the initial mass-radius relation of globular clusters, and we
can only conclude that the present day properties are consistent with
the hypothesis that all hot stellar systems formed with the same
mass-radius relation and that globular clusters have moved away from
this relation because of a Hubble time of stellar and dynamical
evolution.

%
%
\small  
%
%
%
\bibliographystyle{aa}

%
%
%
%


\end{document}